\documentclass{PoS}
\usepackage{gensymb}
\usepackage{graphicx}
\usepackage{cite}
\usepackage[utf8]{inputenc}
\usepackage{url}
\usepackage{placeins}
\usepackage{lineno}
\title{The particle detector in your pocket: The Distributed Electronic Cosmic-ray Observatory}

\ShortTitle{Distributed Electronic Cosmic-ray Observatory}

\author{\speaker{Matthew Meehan}, Silvia Bravo, Felipe Campos, Tyler Ruggles, Cassidy Schneider, Justin Vandenbroucke, Miles Winter\\
         Department of Physics and Wisconsin IceCube Particle Astrophysics Center, University of Wisconsin, Madison, WI 53706, USA\\
        E-mail: \email{mrmeehan@wisc.edu}}
\author{Jeffrey Peacock\\Sensorcast}
\author{Ariel Levi Simons\\Loyola Marymount University}

\abstract{The total area of silicon in cell phone camera sensors worldwide surpasses that in any experiment to date. Based on semiconductor technology similar to that found in modern astronomical telescopes and particle detectors, these sensors can detect ionizing radiation in addition to photons. The Distributed Electronic Cosmic-ray Observatory (DECO) uses the global network of active cell phones in order to detect cosmic rays and other energetic particles such as those produced by radioactive decays. DECO consists of an Android application, database, and public data browser available to citizen scientists around the world (https://wipac.wisc.edu/deco). Candidate cosmic-ray events have been detected on all seven continents and can be categorized by the morphology of their corresponding images. We present the DECO project, a novel particle detector with wide applications in public outreach and education.}

\FullConference{35th International Cosmic Ray Conference --- ICRC2017\\
		10--20 July, 2017\\
		Bexco, Busan, Korea}

\begin{document}

\section{Introduction}\label{sec_intro}
\begin{figure}
  \centering
  \includegraphics[width=.5\linewidth]{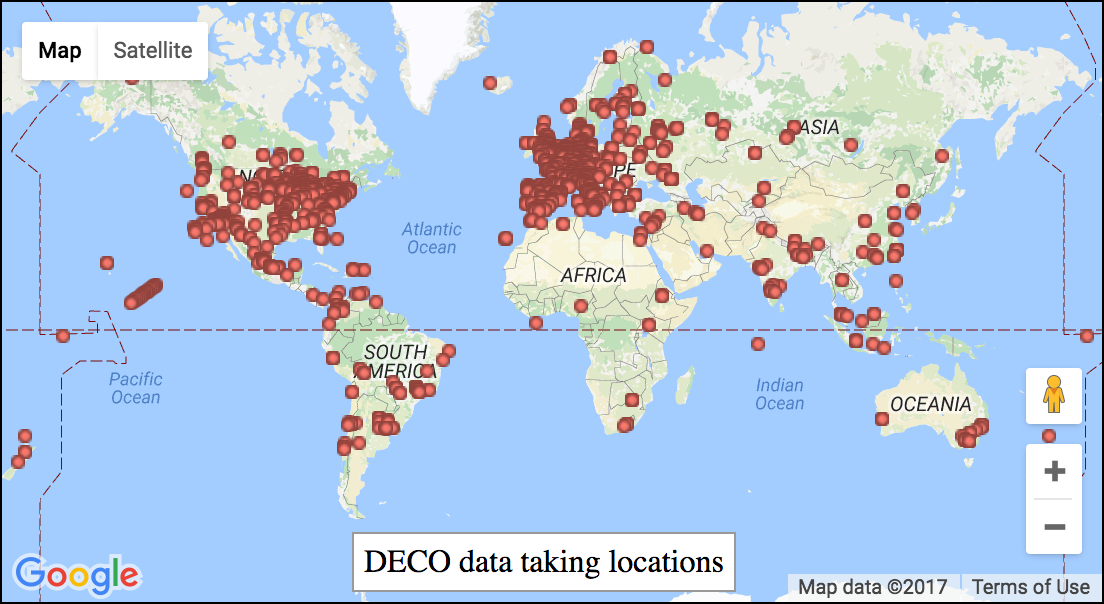}
  \caption{\small{Map of DECO user locations as of June 2017. DECO now has users from 60 different countries on all seven continents (Antarctica not shown).}}
  \label{map}
\end{figure}

The Distributed Electronic Cosmic-ray Observatory (DECO) project is a citizen science project that enables individuals to turn their cell phones into particle detectors~\cite{DECO_paper}. DECO, at its heart, is an Android application that uses smart phone camera sensors to detect ionizing radiation. An iOS version of the app is currently in the beta testing stage of development and expected to be released within the year. The project, which went public in the fall of 2014, has users running on all seven continents (Figure~\ref{map}). DECO and the data it collects are being used for both basic research and pedagogical endeavors, such as introducing students to fundamental concepts in astrophysics, particle physics, computer science, and data analysis. Recently, DECO data was used to measure the thickness of the sensitive region (depletion region) of cell phone image sensors, which is proprietary information from the manufacturer~\cite{deco_depletion_depth}. Currently, a new machine learning analysis is being developed to classify events detected by DECO. This analysis is expected to greatly improve the efficiency of identifying cosmic-ray muons and, in the near future, will give users the ability to select cosmic-ray events for analysis.
% * <mrmeehan@wisc.edu> 2017-06-30T18:10:22.402Z:
% 
% > , and the data it collects,
% Not sure about commas use here.. sound better without it (to me)
% 
% ^ <mrmeehan@wisc.edu> 2017-06-30T18:12:34.431Z.

\section{DECO App}\label{sec_deco_app}
% Describe technology and app algorithm for finding events
The CMOS technology in cell phone camera sensors is similar to that found in professional astronomical telescopes and particle physics experiments. These sensors can detect ionizing radiation via the electron/hole pairs created when ionizing charged particles traverse the depleted region of the sensor. The DECO mobile phone application is designed to be run with the camera lens face down or covered with opaque tape to prevent ambient light from contaminating the images. The app repeatedly takes exposures of \textasciitilde50 ms in duration. Each exposure is run through a filter algorithm that decides if the image contains a candidate charged particle event. The filter has two stages: first a low-resolution image is analyzed and, if passed, a high-resolution image is checked. The filters check for N pixels above a threshold that is determined during a calibration period when the app is first loaded. All images that pass both filters are tagged as candidate events and uploaded to a central database for further analysis. Additionally, one image is saved every 5 minutes from each device in order to gather an unbiased sample of exposures for calibrations and noise studies (minimum bias).

Both candidate events and minimum bias images are stored on a central database for future analysis, which can be browsed at~\cite{deco_data}. The database contains a JPEG image for each event, along with a set of metadata including a unique event ID, unique device ID, phone model, Android version, time, geolocation (rounded to the nearest 0.01$^{\circ}$ in latitude and longitude for privacy), altitude, magnetic field values, pressure, and temperature. The web interface to the database allows users to query based on date, location, device ID, phone model, and event category (standard or minimum bias).

\section{Event Types}\label{sec_events}
\begin{figure}
  \centering
  \includegraphics[width=.8\linewidth]{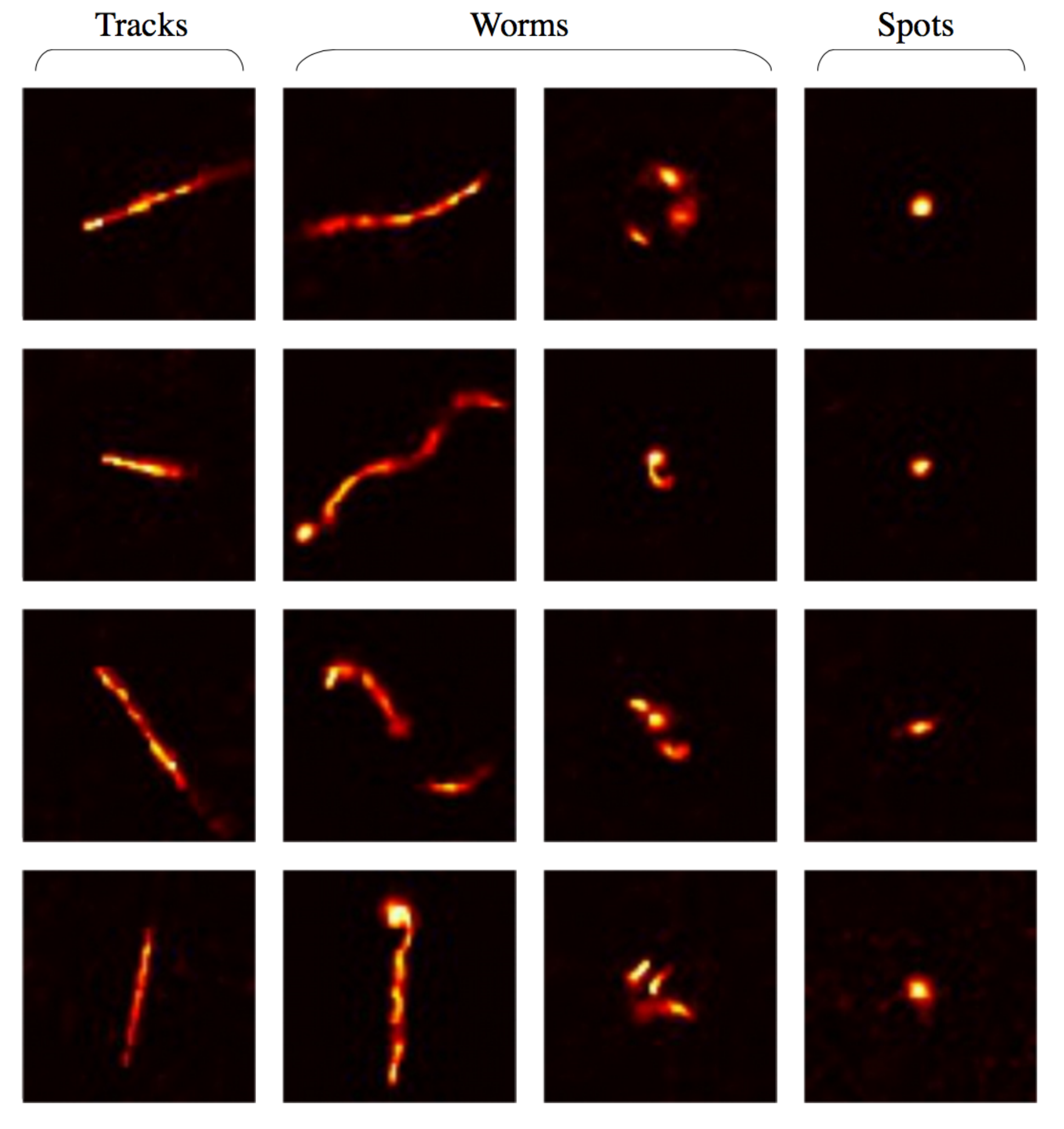}
  \caption{\small{Representative sample of the three distinct types of charged particle events that require classification. Tracks and spots, left and right columns, respectively, are generally observed to have very consistent and predictable features. Worms, middle two columns, are observed to have a much wider variety of features, many of which present potential classification confusion when compared to track-like and spot-like features. Each image above has been converted to grayscale and cropped to dimensions of $64\times64$ pixels.}}
  \label{blob_groups}
\end{figure}

There are three distinct types of charged particle events seen by DECO: tracks, worms, and spots. These are named according to the convention in~\cite{groom}, which categorizes events by the morphology of their resulting images. Tracks are long, straight clusters of bright pixels that are caused by cosmic rays, typically minimum ionizing muons at sea level or primary cosmic rays at altitude greater than several km. Worms are aptly named for their curvy paths, which are produced by low-energy electrons (produced by radioactive decays) undergoing multiple Coulomb scattering interactions inside the sensor.  Spots are smaller, circular clusters of pixels which can be created by lower energy electrons (e.g. produced by Compton scatters of gamma rays from radioactive decay) being quickly absorbed.  Spots could also be produced by alpha particles or by cosmic rays incident normal to the sensor. Figure~\ref{blob_groups} shows the characteristic camera sensor response for each of the three particle physics events seen by DECO. In addition to the particle related event types, there are also events that result from artifacts in the camera image sensor: light exposure, hot spots, thermal noise fluctuations, and larger scale sensor artifacts such as rows of bright pixels \cite{DECO_paper}. 

Given the numerous event types, both particle and non-particle, and the increasing number of images being collected by DECO, there is a growing need for a reliable computerized event classification system. However, there are several challenges associated with characterizing the DECO data set in a way that requires little human intervention. Due to the inhomogeneity in hardware\footnote{Users are running DECO on 604 distinct phone models.} and data acquisition conditions, otherwise identical events may be detected differently, e.g. due to fluctuations in brightness, background noise, number of pixels detecting the event, etc. Additionally, DECO particle events possess high levels of rotational and translational symmetry, all of which can be particularly challenging to account for with traditional model fitting algorithms. Fortunately, advances in the quickly developing field of machine learning offer possible techniques to overcome these modeling challenges. 

In the sections that follow, we describe the preliminary construction and optimization of a DECO-specific convolutional neural network model that can be used to increase both the accuracy and efficiency of DECO event classification. 

\section{Classification With Convolutional Neural Networks}\label{sec_cnn}
Convolutional neural networks (CNNs) are deep learning neural network architectures that have shown extraordinarily good performance on image classification tasks \cite{CNN_paper}. The core concept behind CNNs is to build many layers of ``feature detectors'' that take into account the topological structure of the input data, such as the spatial arrangement of pixels in an image \cite{CNN_best_practices}. Through a process known as training, the model learns how to extract meaningful features from the input, which can then be used to make predictions about the contents of the input. CNNs typically contain two types of alternating layers that are used for ``feature extraction'': convolution layers and pooling (subsampling) layers. 
 
Convolution layers take a stack of inputs (e.g. color channels in an image) and convolve each with a set of learnable filters to produce a stack of output feature maps, where each feature map is simply a filtered version of the input (i.e., an image). Pooling layers reduce the dimensionality of a feature map by computing an aggregation function across small local regions of the input. A common method, known as max pooling, uses an aggregation function that computes the maximum element for each local region of the input. For example, max pooling can be used to reduce a $32\times32$ image to a $16\times16$ image by dividing the image into 256, $2\times2$ grids of pixels and propagating only the maximum pixel value in each $2\times2$ grid to the next layer of the model. As a result of the max pooling operation, only the most pronounced features are forwarded to the subsequent layers of the model. Following convolution and pooling, the feature maps are then used as input for one or more fully connected layers, where classification typically takes place in the last of these layers. A thorough treatment of these topics can be found, for example, in \cite{goodfellow}. 
 
%By exploiting translational symmetry, the combination of convolution and pooling result in restricted connectivity patterns that drastically reduce the number of parameters required to model an image. 

\begin{figure}
  \centering
  \includegraphics[width=1\linewidth]{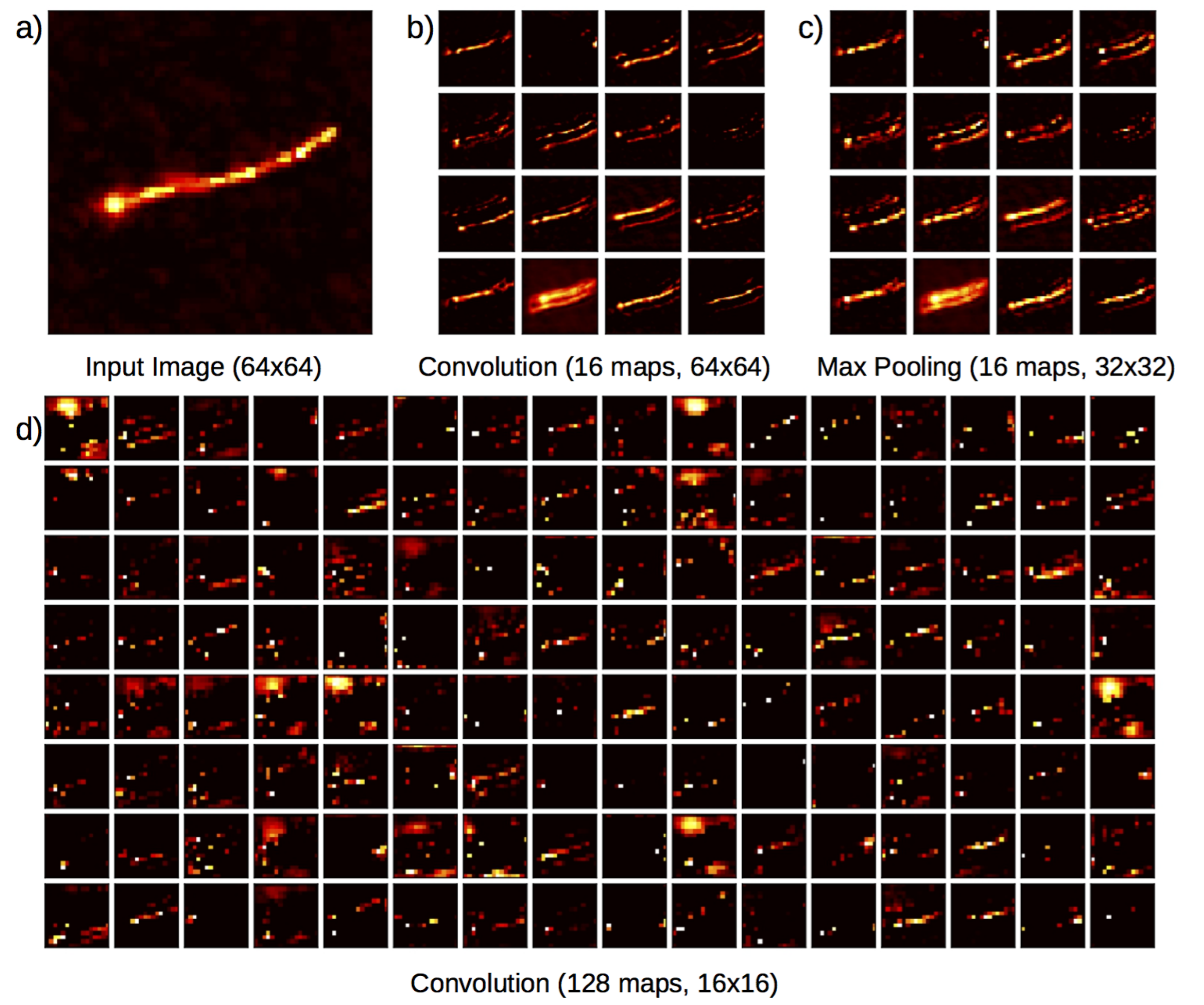}
  \caption{\small{Sample of the layer-by-layer output taken from the layers of the model that perform feature extraction. In each sub-figure, the labels indicate the operation performed, the number of feature maps, and the output dimensions of each feature map. (a) Input image after being converted to grayscale and cropped to $64\times64$ pixels. (b) Output feature maps created by convolving the input image with a set of 16 learned filters. (c) Output after performing subsampling (in this case, max pooling, as explained in Section~\ref{sec_cnn}) on the output feature maps shown in (b). (d) Output feature maps after performing three more convolutions and one additional subsampling (max pooling) operation on the feature maps shown in (c).}}
  \label{cnn_layers}
\end{figure}

\subsection{Image Data Set and CNN Training}\label{sec_training}
The initial image data set consisted of approximately 1000 images, each of which potentially contained the signature of a charged particle event. The images were then inspected by eye, by multiple individuals, and assigned labels of track, spot, worm, ambiguous, or other. If an image contained noise artifacts, such as a hot spot in the camera sensor, the image was discarded. Additionally, if a clear identification could not be made, the image was labeled as ambiguous and excluded from the image set. The remaining images (spots, worms, and tracks), along with their labels, were then used to generate a database containing approximately 700 images, 600 for CNN training and 100 for testing the model. Each image in the database was converted to grayscale and cropped to dimensions of $64\times64$ pixels (as shown for example images in Figure~\ref{blob_groups}). Grayscale (unweighted sum of red, green, and blue value of each pixel) was chosen to reduce the variation observed from device to device. At the beginning of each training epoch (iteration of 600 images), the images were given a series of random perturbations consisting of horizontal and vertical reflections, zoom in or out up to 10$\%$, horizontal and vertical shifts, and rotations between $0^\circ$ and $360^\circ$. The perturbations are used to build in rotational invariance and to ensure that the model never sees the exact same version of an image more than once. 

The CNN used for this study was constructed in Python and made use of the Theano Python library \cite{theano} and the Keras neural network API \cite{keras}. The best performing model consists of eight convolution layers, with $2\times2$ max pooling performed after every second convolution layer, followed by three fully connected layers for classification. Figure~\ref{cnn_layers} shows the output of select layers in the convolution and pooling section of the CNN. 
%Additional procedures to describe if space permits: how training works, dropout, class weighting, label smoothing, training length... 

\subsection{Preliminary Classification Results}\label{cnn_results}
\begin{figure}
  \centering
  \includegraphics[width=1\linewidth]{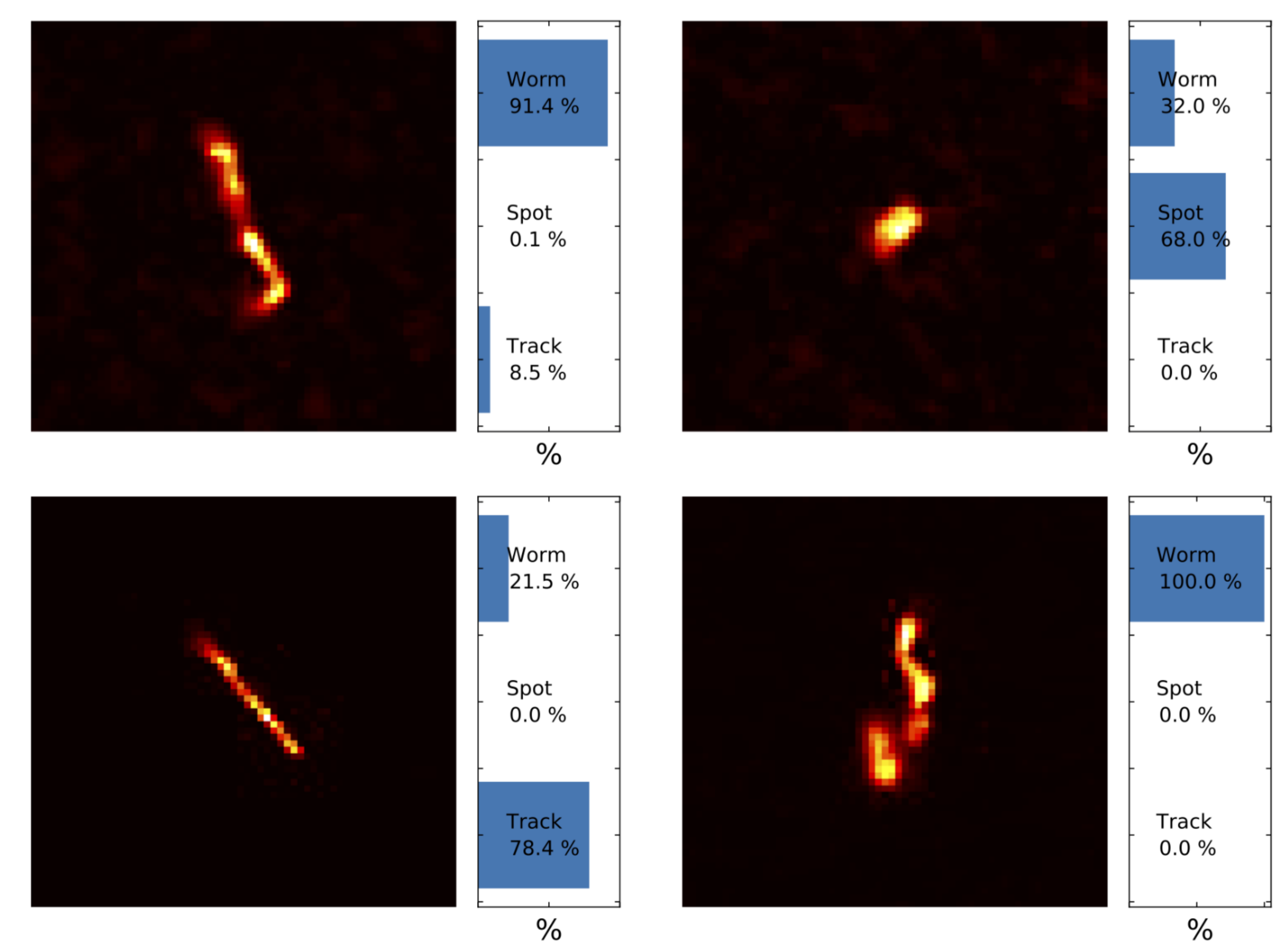}
  \caption{\small{Several charged particle events and their corresponding classifications. For each event, the likelihood of belonging to each class, as determined by the CNN, is shown. }}
  \label{predictions}
\end{figure}

The best performing model contained over 14 million learnable parameters and was trained on a GPU for a total of 6000 training epochs, which corresponds to 3.6 millions images being seen by the model. Following training, the model was used to make predictions about the correct classification of new images. The trained model accepts an input image whose classification is unknown and outputs a list of probabilities, one for each of the three classes. An example of this process is shown in Figure~\ref{predictions}. The class with the highest probability is taken to be the CNN classification. 

\begin{figure}
  \centering
  \includegraphics[width=1\linewidth]{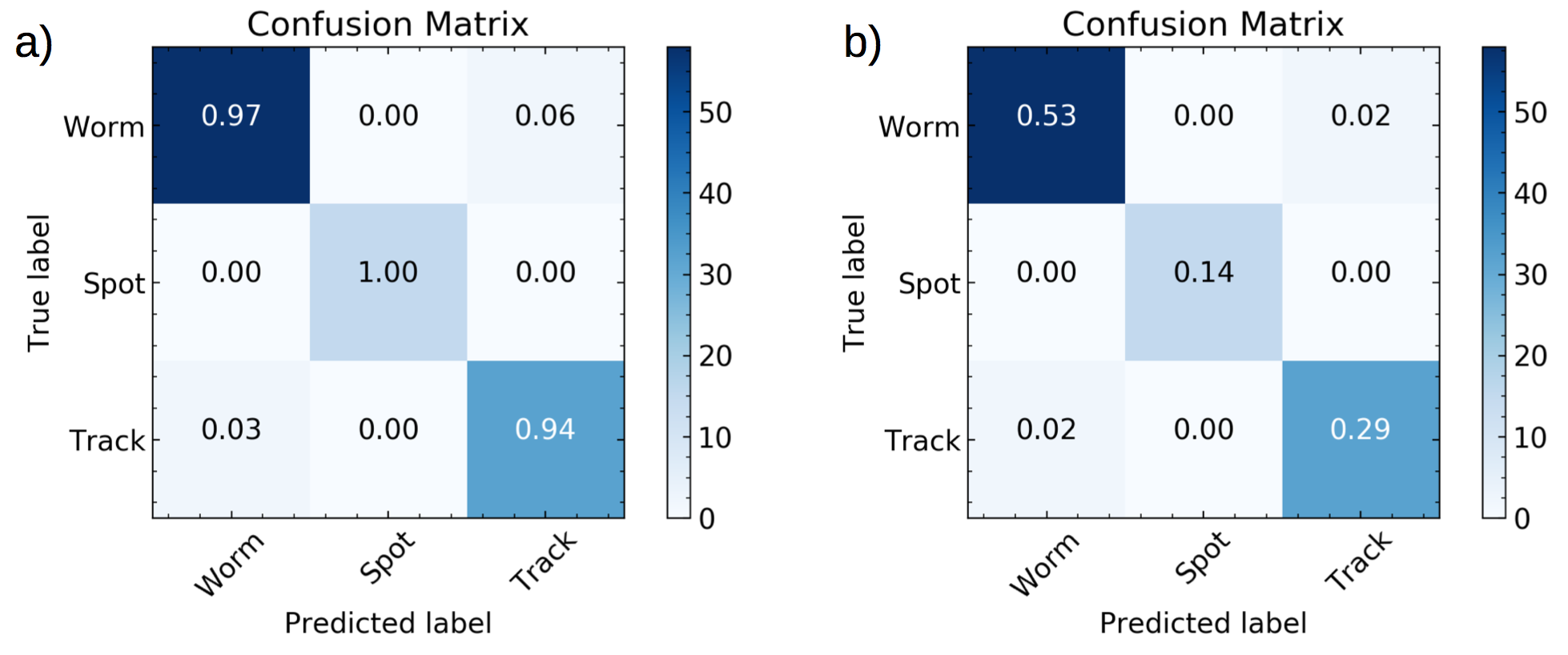}
  \caption{\small{Confusion matrices summarizing the accuracy of results based on an evaluation of 109 images. (a) Normalized confusion matrix where each entry is shown as a percentage calculated from the total number of entries in each column. (b) Confusion matrix presented as a 2D PDF showing the percentage for each category, normalized to the total number of events. For both plots, the vertical axis shows the ground truth (human-determined) classification and the horizontal axis shows the classification determined by the best-performing computer classification. The color axis represents the number of images in each category.}}
  \label{conf_mat}
\end{figure}

To estimate the overall accuracy of the model, an independent testing set of images and labels were evaluated. The evaluation set consisted of approximately 100 images that were not involved in the training of the model. The predictions of the model were then tested against the human-assigned labels for each image. After evaluating each image, the accuracy of the model was found to be approximately $96\%$. Figure~\ref{conf_mat} shows a class-by-class summary, known as a confusion matrix, quantifying the model accuracy. The majority of incorrect classifications are tracks that are falsely assigned the classification of worm. 
 
%\begin{figure}
%  \centering
%  \includegraphics[width=1\linewidth]{./plots/best_cm_gpu_vertical_norm.png}
%  \caption{\small{Vertically normalized confusion matrix.  Entries are shown as percentages. The vertical axis shows the ground truth (human-determined) classification.  The horizontal axis shows the classification determined by the best-performing computer classification. The color axis represents the number of images in each category.}}
%  \label{conf_mat}
%\end{figure}

\section{Conclusion}\label{sec_conclusion}
The classification analysis presented is ongoing and the results are preliminary. The initial results are encouraging and the CNN classification will eventually be included in the regular processing of the data. The public database will include this classification, allowing any user to select either cosmic-ray or radioactivity-induced events from the data set with high confidence. The biggest limitation to improving the current model is the size of the event sample used for training, which needs to encompass the variation in particle physics event types and sensor response over hundreds of different phone models. This is constantly improving as the size of the DECO user base increases. An iOS version of DECO is currently under development and expected to be released within the next year. There are far fewer distinct models of smart phones running iOS compared to Android, resulting in a more homogeneous data set. We expect the the classification analysis to be more efficient on the iOS data set.  Reliable event classification available to users will help expand the worldwide reach of DECO.  

\section{Acknowledgements}\label{sec_acknowledgments}
DECO is supported by the American Physical Society, the Knight Foundation, the Simon Strauss Foundation, and QuarkNet. We are grateful for beta testing, software development, and valuable conversations with Keith Bechtol, Segev BenZvi, Andy Biewer, Paul Brink, Patricia Burchat, Duncan Carlsmith, Alex Drlica-Wagner, Mike Duvernois, Lucy Fortson, Stefan Funk, Mandeep Gill, Laura Gladstone, Giorgio Gratta, Jim Haugen, Kenny Jensen, Kyle Jero, David Kirkby, David Saltzberg, Marcos Santander, and Ian Wisher.

% FloatBarrier forces floating figures to be above references, but then the 8 page limit is exceeded. We can probably adjust image sizes until it all fits and looks nice.

%\FloatBarrier
\bibliography{DECO_ICRC_2017}{}
\bibliographystyle{plain}
\end{document}